\documentclass[10pt]{bmc_article}    

\usepackage{cite} 
\usepackage{url}  
\usepackage{ifthen}  
\usepackage{multicol}   
\usepackage[latin1]{inputenc} 
\usepackage{mathbbol,amssymb,latexsym,amsfonts,amsmath,amsthm}
\usepackage{graphicx}
\usepackage{float}
\usepackage[vlined]{algorithm2e}

\newcommand{\lemone}{\textsf{LeMoNe}}
\newcommand{\syntren}{SynTReN}

\newcommand{\etal}{\textit{et al.}{ }}
\newcommand{\Pa}{\mathrm{Pa}}
\newcommand{\A}{\mathcal{A}}

\newcommand{\E}{\mathcal{E}}
\newcommand{\R}{\mathcal{R}}
\newcommand{\prs}{P}
\newcommand{\rec}{R}
\newcommand{\tp}{\mathrm{tp}}
\newcommand{\fp}{\mathrm{fp}}
\newcommand{\fn}{\mathrm{fn}}
\renewcommand{\P}{\mathcal{P}}
\newcommand{\iintsp}{\iint\hspace*{-3pt}}

\usepackage{fancyhdr}

\newboolean{publ}

\newenvironment{bmcformat}{\fussy\setboolean{publ}{true}}{\fussy}

\begin{document}
\begin{bmcformat}


\vspace*{-6em}

\title{Validating module network learning algorithms using simulated data}

\author{Tom~Michoel\correspondingauthor$^{\natural 1}$%
  \email{\correspondingauthor tom.michoel@psb.ugent.be}%
  \and  Steven Maere$^{\natural 1}$\email{steven.maere@psb.ugent.be}%
  \and Eric Bonnet$^{1}$\email{eric.bonnet@psb.ugent.be}%
  \and Anagha Joshi$^{1}$\email{anagha.joshi@psb.ugent.be}%
  \and Yvan Saeys$^{1}$\email{yvan.saeys@psb.ugent.be}%
  \and Tim Van den Bulcke$^2$\email{tim.vandenbulcke@esat.kuleuven.ac.be}%
  \and  Koenraad Van Leemput$^3$\email{koen.vanleemput@ua.ac.be}%
  \and Piet van Remortel$^3$\email{piet.vanremortel@ua.ac.be}%
  \and Martin Kuiper$^1$\email{martin.kuiper@psb.ugent.be}%
  \and  Kathleen Marchal$^{2,4}$\email{kathleen.marchal@biw.kuleuven.be}%
  { }and  Yves Van de Peer$^1$\email{yves.vandepeer@psb.ugent.be}}


\address{%
  \iid(1)Bioinformatics \& Evolutionary Genomics, Department of Plant
  Systems Biology,  VIB/Ghent University, Technologiepark 927, B-9052 Ghent, Belgium\\
  \iid(2)ESAT-SCD, K.U.Leuven, Kasteelpark Arenberg 10, B-3001 Leuven, Belgium\\
  \iid(3)ISLab, Department of Mathematics and Computer Science, University of Antwerp,
  Middelheimlaan 1, B-2020 Antwerpen, Belgium\\
  \iid(4)CMPG, Department Microbial and Molecular Systems, K.U.Leuven, Kasteelpark Arenberg 20,
  B-3001 Leuven, Belgium
}%

\maketitle

\noindent{ }\textsf{{\footnotesize $^\natural$Contributed equally}} 


\begin{abstract}
  \paragraph*{Background:} 
  In recent years, several authors have used probabilistic graphical
  models to learn expression modules and their regulatory programs
  from gene expression data. Despite the demonstrated success of such
  algorithms in uncovering biologically relevant regulatory relations,
  further developments in the area are hampered by a lack of tools to
  compare the performance of alternative module network learning
  strategies. Here, we demonstrate the use of the synthetic data
  generator \syntren{} for the purpose of testing and comparing module
  network learning algorithms. We introduce a software package for
  learning module networks, called \lemone{}, which incorporates a
  novel strategy for learning regulatory programs. Novelties include
  the use of a bottom-up Bayesian hierarchical clustering to construct
  the regulatory programs, and the use of a conditional entropy
  measure to assign regulators to the regulation program nodes. Using
  \syntren{} data, we test the performance of \lemone{} in a
  completely controlled situation and assess the effect of the
  methodological changes we made with respect to an existing software
  package, namely Genomica. Additionally, we assess the effect of
  various parameters, such as the size of the data set and the amount
  of noise, on the inference performance.

  \paragraph*{Results:} 
  Overall, application of Genomica and \lemone{} to simulated data sets
  gave comparable results. However, \lemone{} offers some advantages,
  one of them being that the learning process is considerably faster
  for larger data sets. Additionally, we show that the location of the
  regulators in the \lemone{} regulation programs and their
  conditional entropy may be used to prioritize regulators for
  functional validation, and that the combination of the bottom-up
  clustering strategy with the conditional entropy-based assignment of
  regulators improves the handling of missing or hidden regulators.
  
  \paragraph*{Conclusions:} 

  We show that data simulators such as \syntren{} are very well suited
  for the purpose of developing, testing and improving module network
  algorithms. We used \syntren{} data to develop and test an
  alternative module network learning strategy, which is incorporated
  in the software package \lemone{}, and we provide evidence that this
  alternative strategy has several advantages with respect to existing
  methods.
\end{abstract}

\ifthenelse{\boolean{publ}}{\begin{multicols}{2}}{}


\section*{Background}

For the past 45 years, research in molecular biology has been based
predominantly on reductionist thinking, trying to unravel the complex
workings of living organisms by investigating genes or
proteins one at a time. In recent years, molecular biologists have
come to view the cell from a different, more global perspective. With
the advent of fully sequenced genomes and high-throughput functional
genomics technologies, it has become possible to monitor molecular
properties such as gene expression levels or protein-DNA interactions
across thousands of genes simultaneously. As a consequence, it has
become feasible to study genes, proteins and their interactions in the
context of biological systems rather than in isolation. This novel
paradigm has been named `systems biology' \cite{idek01b}.

One of the goals of the systems approach to molecular biology is to
reverse engineer the regulatory networks underlying cell function.
Particularly transcriptional regulatory networks have received a lot
of attention, mainly because of the availability of large amounts of
relevant experimental data. Several studies use expression data,
promoter motif data, chromatin immunoprecipitation (ChIP) data and/or
prior functional information (e.g. GO classifications \cite{ashb00} or
known regulatory network structures) in conjunction to elucidate
transcriptional regulatory networks \cite{frie00, peer01,
  barj03, segal2003, beer2004, friedman2004, harbison2004,
  luscombe2004, xu2004, batt05, basso2005, gart05, pett05, lemm06,
  vandenbulcke2006}. Most of these methods try to unravel the
\emph{control logic} underlying specific expression patterns. 
This type of analysis typically requires elaborate computational
frameworks. In particular probabilistic graphical models are
considered a natural mathematical framework for inferring regulatory
networks \cite{friedman2004}. Probabilistic graphical models, the
best-known representatives being Bayesian networks, represent the
system under study in terms of conditional probability distributions
describing the observations for each of the variables (genes) as a
function of a limited number of parent variables (regulators), thereby
reconstructing the regulatory network underlying the observations.
Friedman \emph{et al.} pioneered the use of Bayesian networks to learn
regulatory networks from expression data \cite{frie00, peer01}. In
these early studies, each gene in the resulting Bayesian network is
associated with its individual regulation program, i.e., its own set
of parents and conditional probability distribution. A key limitation
of this approach is that a vast number of structural features and
distribution parameters need to be learned given only a limited number
of expression profiles. In other words, the problem of finding back
the real network structure is typically heavily underdetermined. An
attractive way to remedy this issue is to take advantage of the
inherent modularity of biological networks \cite{hartwell1999},
specifically the fact that groups of genes acting in concert are often
regulated by the same regulators. Segal \emph{et al.}
\cite{segal2003,segal2005b} first exploited this idea by proposing
\emph{module networks} as a mathematical model for regulatory
networks. Module networks are probabilistic graphical models in which
groups of genes, called \emph{modules}, share the same parents and
conditional distributions. As the number of parameters to be estimated
in a module network is much smaller than in a full Bayesian network,
the currently available gene expression data sets can be large enough
for the purpose of learning module networks \cite{segal2003, xu2004,
  batt05, segal2005b}.

Despite the demonstrated success of module network learning algorithms
in finding biologically relevant regulatory relations \cite{segal2003,
  xu2004, batt05, segal2005b}, there is only limited information about
the actual recall and precision of such algorithms
\cite{batt05} and how these performance measures are influenced by the
use of alternative module network learning strategies. Having the
means to answer the latter question is key to the further development
and improvement of the module networks formalism. 

The purpose of the present study is twofold. First, we introduce a
novel software package for learning module networks, called \lemone{},
which is based on the general methodology outlined in Segal {\it et
  al.} \cite{segal2003} but incorporates an alternative strategy for
inferring regulation programs.

Second, we demonstrate the use of \syntren{} \cite{vandenbulcke2006b},
a data simulator that creates synthetic regulatory networks and
produces simulated gene expression data, for the purpose of testing
and comparing module network learning algorithms. We use \syntren{}
data to assess the performance of \lemone{} and to compare the
behavior of alternative module network learning strategies.
Additionally, we assess the effect of various parameters, such as the
size of the data set and the amount of noise, on the inference
performance. For comparison, we also use \lemone{} to analyze real
expression data for \emph{S. cerevisiae} \cite{gasch2000} and
investigate to what extent the quality of the module networks learned
on real data can be automatically assessed using structured biological
information such as GO information and ChIP-chip data
\cite{harbison2004}.

\section*{Methods}

\subsection*{Data sets}

We used \syntren{} \cite{vandenbulcke2006b} to generate simulated data
sets for a gene network with 1000 genes of which 105 act as
regulators. The topology of the network is subsampled from an \emph{E.
  coli} transcriptional network \cite{ma2004} by cluster addition,
resulting in a network with $2361$ edges. All parameters of \syntren{}
were set to default values, except number of correlated inputs, which
was set to 50\%.  \syntren{} generated expression values ranging from
$0$ (no expression) to $1$ (maximal expression) which we normalized to
$\log_2$ ratio values by picking one of the experiments as the
control.  Except where indicated otherwise, the list of true
regulators was given as the list of potential regulators for \lemone{}
and Genomica.

For the tests performed on real data, we used an expression
compendium for {\it S. cerevisiae} containing expression data for 173
different experimental stress conditions \cite{gasch2000}. The data were
obtained in prenormalized and preprocessed form. We used the mean
$\log_{2}$ values of the expression ratios (perturbation vs. control).

To assess the quality of the regulatory programs learned from real
data, we used data on genome-wide binding and phylogenetically
conserved motifs for 102 transcription factors from Harbison \emph{et
  al.} \cite{harbison2004}. For a given transcription factor, only
genes that were bound with high confidence (significance level $\alpha
= 0.005$) and showed motif conservation in at least one other {\it
  Saccharomyces} species (besides {\it S. cerevisiae}) were considered
true targets.

\subsection*{Module networks}

Module networks are a special kind of Bayesian networks and were
introduced by Segal \etal \cite{segal2003,segal2005}. To each gene $i$
we associate a random variable $X_i$ which can take continuous values
and corresponds to the gene's expression level. The distribution of
$X_i$ depends on the expression level of a set of parent genes $\Pa_i$
chosen from a list of potential regulators. If the network formed by
drawing directed edges from parent genes to children genes is acyclic,
we can define a joint probability distribution for the expression
levels of all genes as a product of conditional distributions,
\begin{equation}\label{eq:7}
  p(x_1,\dots,x_N) = \prod_{i=1}^N p_i \bigl(x_i \mid \{x_j\colon 
  j\in\Pa_i\}\bigr).
\end{equation}
This is the standard Bayesian network formalism. 

In a module network we assume that genes are partitioned into
different sets called \emph{modules}, such that genes in the same
module share the same parameters in the distribution function
(\ref{eq:7}). Hence a module network is defined by a partition of
$\{1,\dots,N\}$ into $K\ll N$ modules $\A_k$ such that
$\cup_{k=1}^K\A_k = \{1,\dots,N\}$ and $\A_k\cap\A_{k'}=\emptyset$ for
$k\neq k'$, a collection of parent genes $\Pi_k$ for each module $k$,
and a joint probability distribution
\begin{equation}\label{eq:2}
  p(x_1,\dots,x_N) = \prod_{k=1}^K \prod_{i\in\A_k} p_k 
  \bigl(x_i \mid \{x_j\colon   j\in\Pi_k\}\bigr).
\end{equation}
The conditional distribution $p_k$ of the expression level of the
genes in module $k$ is normal with mean and standard deviation
depending on the expression values of the parents of the module
through a regression tree that is called the \emph{regulation program}
of the module. The tests on the internal nodes of the regression tree
are of the form $x\gtrless s$ for some split value $s$, where $x$ is
the expression value of the parent associated to the node (Figure
\ref{fig:gasch_module_11}).

The Bayesian score is obtained by taking the log of the marginal
probability of the data likelihood over the parameters of the normal
distributions at the leaves of the regression trees with a
normal-gamma prior (see \cite{segal2005} and Additional file 1 for
more details; the actual expression for the score is in
eq.~(\ref{eq:S9})).  Its main property is that it decomposes as a sum
of leaf scores of the different modules:
\begin{align}
  \mathbf{S} &= \sum_k \mathbf{S}_k = \sum_k\sum_\ell S_k(\E_\ell),
  \label{eq:12}
\end{align}
where $\E_\ell$ denotes the experiments that end up at leaf $\ell$
after traversing the regression tree. A normal-gamma prior ensures
that $S_k(\E_\ell)$ can be solved explicitly as a function of the
sufficient statistics (number of data points, mean and standard
deviation) of the leaves of the regression tree (see Additional file
1).

\subsection*{Learning module regulation programs}

For a given assignment of genes to modules, finding a maximum for the
Bayesian score (\ref{eq:12}) consists of finding the optimal
partitioning of experiments into `leaves' $\ell$ for each module
separately, i.e., find a collection of subsets
$\E_\ell\subset\{1,\dots,M\}$ such that $\cup_\ell \E_\ell =
\{1,\dots,M\}$, $\E_{\ell}\cap\E_{\ell'}=\emptyset$ for
$\ell\neq\ell'$, and
\begin{equation}\label{eq:13}
  \mathbf{S}_k = \sum_\ell S_k(\E_\ell)
\end{equation}
is maximal. In particular we do not have to define the parent sets
$\Pi_k$ of the modules in order to find an optimal partition.

We use a bottom-up hierarchical clustering method to heuristically
find a high-scoring partition.  At each step of the process we have a
collection of binary trees $T_\alpha$ which represent subsets
$\E_\alpha$ of experiments. The binary split of $T_{\alpha}$ into its
children $T_{\alpha_1}$ and $T_{\alpha_2}$ corresponds to a partition
of the set $\E_{\alpha}$ into two sets:
$\E_{\alpha}=\E_{\alpha_1}\cup\E_{\alpha_2}$. The initial collection
consists of trivial trees without children representing single
experiments.  To proceed from one collection of trees to the next, the
pair of trees with highest merge score is merged into a new tree, and
the collection of binary trees decreases by one, eventually leading to
one hierarchical tree $T_0$ representing the complete experiment set
$\E_0=\{1,\dots,M\}$. The simplest merge score is given by the
possible gain in Bayesian score by merging two experiment sets:
\begin{equation}\label{eq:11}
  r_{\alpha_1,\alpha_2} = S_k(\E_{\alpha_1}\cup\E_{\alpha_2}) 
  - S_k(\E_{\alpha_1}) - S_k(\E_{\alpha_2}).
\end{equation}
In Additional file 1 we define an alternative merge score related to
the Bayesian hierarchical clustering method of \cite{heller2005}. This
merge score takes into account the substructure of the trees below
$T_{\alpha_1}$ and $T_{\alpha_2}$ in addition to the Bayesian score
difference (\ref{eq:11}), and tends to produce more balanced trees.

In the final step, we need to cut the hierarchical tree $T_0$. To this
end we traverse the tree from the root towards its leaves. If we are
at a subtree node $T_\alpha$ with children $T_{\alpha_1}$ and
$T_{\alpha_2}$, we compute the score difference (\ref{eq:11}). If this
difference is negative, the total score is improved by keeping the
split $T_\alpha$, and we move on to test each of its children nodes.
If the difference is positive, the total score is improved by not
making the split $T_\alpha$, and we remove its children nodes from the
tree. The experiment set $\E_\alpha$ becomes one of the leaves of the
regulation program, contributing one term in the sum (\ref{eq:13}).

The pseudocode for the regulation program learning algorithm is given
in Figure \ref{fig:code} in Additional file 1.

In \cite{segal2003,segal2005}, regulation programs are learned
top-down by considering all possible splits on all current leaves with
all potential regulators, so regulation trees and regulator
assignments are learned simultaneously. As a result missing regulators
or noise in the regulator data might lead to a suboptimal partitioning
of the experiments in a module. In our approach we have focused on
finding an optimal partition of the module regardless of the set of
potential regulators. A module collects the data of many genes and
therefore this partition will be less affected by noise or missing
data than when it is determined by exact splits on single regulators.

\subsection*{Regulator assignment}
\label{sec:regulator-assignment}

At a given internal node $T_\alpha$ of the regulation tree $T_0$, the
experiment set $\E_\alpha$ is partitioned into two distinct sets
$\E_{\alpha_1}$ and $\E_{\alpha_2}$ according to the tree structure.
Given a regulator $r$ and split value $s$, we can also partition
$\E_\alpha$ into two sets
\begin{align*}
  \R_1 &= \{m\in\E_\alpha\colon x_{r,m}\leq s \}\\
  \R_2 &= \{m\in\E_\alpha\colon x_{r,m}> s \},
\end{align*}
where $x_{r,m}$ is the expression value of regulator $r$ in experiment
$m$.  

Consider now two random variables: $E$ which can take the values
$\alpha_1$ or $\alpha_2$, and $R$ which can take the values $1$ or
$2$, with probabilities defined by simple counting,
$p(E=\alpha_1)=|\E_{\alpha_1}|/|\E_\alpha|$,
$p(R=1)=|\R_1|/|\E_\alpha|$, etc.  We are interested in the
uncertainty in $E$ given knowledge (through the data) of $R$, i.e., in
the conditional entropy \cite{shannon1948}
\begin{equation}\label{eq:6}
  H(E\mid R) = p_1 h(q_1)  + p_2 h(q_2),
\end{equation}
where $p_i=p(R=i)$, $h$ is the binary entropy function
\begin{equation*}
  h(q) = -q\log(q)-(1-q)\log(1-q),
\end{equation*}
and $q_i$ are the conditional probabilities
\begin{equation*}
  q_i=p(E=\alpha_1\mid R=i)=\frac{|\E_{\alpha_1}\cap \R_i|}{|\R_i|},
  \; i=1,2.
\end{equation*}
In the presence of missing data, the probabilities $p_i$ and $q_i$
need to be modified to take into account this extra uncertainty,
details are given in Additional file 1.

The conditional entropy is nonnegative and reaches its minimum value
$0$ when $q_1=0$ or $1$ (and consequently $q_2=1$, resp. $0$), which
means the $\E$ and $\R$ partitions are equal and the regulator --
split value pair `explains' the split in the regulation tree exactly.
Hence we assign to each internal node of a regulation tree the
regulator -- split value pair which minimizes the conditional
entropy~(\ref{eq:6}).  Since this assignment has to be done only once,
after the module networks score has converged, the best regulator --
split value pairs can be found by simply enumerating over all
possibilities, even for relatively large data sets.  The actual
algorithm for assigning regulators to all nodes operates first on
nodes closer to the roots of the trees where the most significant
splits are located, and takes into account acyclicity constraints on
the module network. It is presented in pseudocode in Figure
\ref{fig:code2} in Additional file 1.

\subsection*{Learning module networks}

To find an optimal module network, learning of regulation trees is
alternated with reassigning genes to other modules until convergence
of the Bayesian score. Module initialization can be done using any
clustering algorithm. Here, we used $k$-means \cite{hoon2004}, and
reassigning is done like in \cite{segal2005} by making all single-gene
moves from one module to another which improve the total score.

\subsection*{Network comparison}

To obtain a gene network from a module network, we put directed edges
from the regulators of a module to all the genes in that module. We
compare inferred to true network by computing the number of edges that
are true positive ($\tp$), false positive ($\fp$) and false negative
($\fn$). Standard measures for the inference quality are precision and
recall. Precision (denoted $\prs$) is defined as the fraction of
edges in the inferred module network that is correct, and recall
(denoted $\rec$) as the fraction of edges in the true network that is
correctly inferred, i.e.,
\begin{align*}
  \prs = \frac{\tp}{\tp + \fp}
  && \rec = \frac{\tp}{\tp + \fn}.
\end{align*}
The $F$-measure, defined as the harmonic mean of precision and
recall, $F = \frac{2\prs \rec}{\prs+\rec}$,
can be used as a single measure for inference quality.

The module content for different module networks can be compared by
computing for each module in one network how many genes of it are also
grouped together in one module in the other network, and averaging
over the number of modules. We call this the average module overlap.

\subsection*{GO overrepresentation analysis}

GO enrichment $P$-values for all modules were determined using the
BiNGO tool \cite{maer05b}, which was incorporated into the \lemone{}
package. The overrepresentation of GO Biological Process categories
was tested using hypergeometric tests and the resulting $P$-values
were corrected for multiple testing using a False Discovery Rate
correction.

\subsection*{Software}

The latest version of \syntren{} can be downloaded from \cite{syntren}
and the latest version of Genomica from \cite{genomica}. \lemone{} is
implemented in Java and available for download in source or executable
form \cite{lemone}.

\section*{Results and Discussion}

\subsection*{Implementation differences in \lemone{} versus Genomica}

As a starting point for the development of \lemone{}, we
re-implemented the methodology described by Segal {\it et al.}
\cite{segal2003}, which is incorporated in the Genomica software
package. Briefly, Genomica takes as input a gene expression data set
and a list of potential regulators. After an initial clustering step,
the algorithm iteratively constructs a regulatory program for each of
the modules (clusters) in the form of a regression tree, and then
reassigns each gene to the module whose program best predicts the
gene's expression behavior. These two steps are repeated until
convergence is reached. In this process, the algorithm attempts to
maximize a Bayesian score function that evaluates the model's fit to
the data \cite{segal2003}.

We used the same overall strategy and the same Bayesian score function
in \lemone{}. However, with respect to the original methods described
by Segal {\it et al.} \cite{segal2003}, \lemone{} incorporates an
alternative strategy for inferring regulatory programs that offers
some advantages (see Methods). First, \lemone{} uses a Bayesian
hierarchical clustering strategy to learn the regulation trees for the
modules from the bottom up instead of from the top down. Furthermore,
contrary to Genomica \cite{segal2003}, the partitioning of expression
data inside a module is not dependent on the expression profiles of
the potential regulators, but only on the module data itself. This
should allow the program to better handle missing or `hidden'
regulators (see further). As an additional advantage, the assignment
of regulators to regulation program nodes can be postponed until after
the final convergence of the Bayesian score, which leads to
considerable time savings (see further).

A second modification in \lemone{} is that regulators are assigned to
the splits in the regulation tree (data splits) based on an
information theoretic measure, namely the conditional entropy of the
partition of the regulator's expression profile dictated by the data
split, given the partition imposed by a particular split value (see
Methods). As a consequence, a data split does not impose, but merely
prefers, a clean partition of the best-matching regulator's expression
values around a certain split value. In comparison with Genomica,
where only such clean partitions are used, this strategy has the
advantage that potential noise in the regulator's expression is taken
into account. Additionally, the conditional entropy can be used to
estimate the quality of the regulator assignment, and thus suggest
missing potential regulators for splits without a low-entropy
regulator. Information theory has been used before to analyze and
cluster gene expression data \cite{butte2000, butte2000b, peer2002,
  sinkkonen2002, kasturi2003, basso2005}. Our method introduces
elements of information theory into the module networks formalism.

In the following sections, we use \syntren{} data to test \lemone{} in
a completely controlled situation in which simulated microarray data
is analyzed for a known underlying regulatory network of reasonable
size, and we assess the performance effects of the aforementioned
methodological changes with respect to Genomica \cite{segal2003}. The
\lemone{} package and the source code are freely available under the
GPL license (see Software section).

\subsection*{Modularity}

A fundamental assumption of the module networks formalism is that real
biological networks have a modular structure \cite{hartwell1999} that
is reflected in the gene expression data, and therefore groups of
genes can share the same parameters in the mathematical description of
the network. In \lemone{}, as in other module network learning
programs \cite{segal2003, xu2004}, the desired number of modules has
to be given as an input parameter to the inference program, and a main
question is how the optimal module number has to be determined. Fewer
modules means lower computational cost and more data points per
module. This results in a better estimation of parameters, but
possibly entails oversimplifying the network and missing important
regulatory relations. More modules means more specific optimization of
the network at higher computational cost. When modules become too
small, there could be too few data points per module for a reliable
estimation of the parameters. In this section we use the Bayesian
score to estimate the optimal number of modules. 

\begin{figure}[H]
  \centering
  \includegraphics[width=\linewidth]{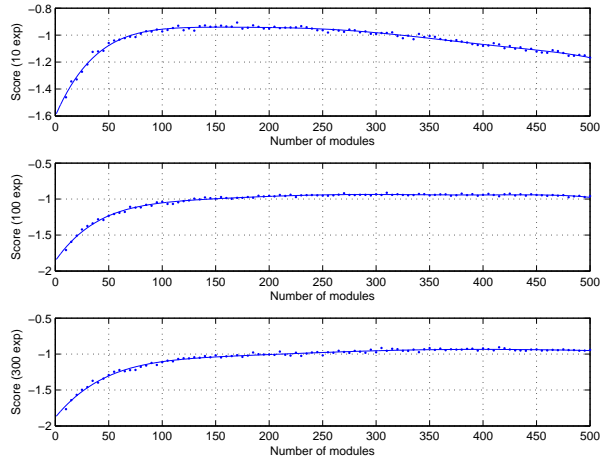}
  \caption{Bayesian score as a function of the number of modules for
    data sets with 10, 100 and 300 experiments (top to bottom). The
    score is normalized by the number of genes times the number of
    experiments. The curves are least squares fits of the data to a
    linear non-polynomial model of the form $a_0 + \sum_{k=1}^n a_k
    x^{k-1} e^{-x/500}$ with $x$ the number of modules and $n=6$.}
  \label{fig:score_mod}
\end{figure}

Throughout this manuscript, we make use of a \syntren{}-generated
synthetic network encompassing 1000 genes of which 105 act as
regulators (see Methods). Unless otherwise stated, we use all 105
regulators in this network as potential regulators while inferring
module networks.  Figure~\ref{fig:score_mod} shows the Bayesian score,
normalized by the number of genes times the number of experiments, for
this network and different numbers of experiments. In all three
panels, the score reaches a maximum.  The top panel (data set with 10
experiments), which has a true maximum for the score, illustrates that
the network inference problem is underdetermined for very small data
sets. Increasing the number of modules beyond the location of the
maximum lowers the fit of the model to the data.  For larger data sets
(middle and bottom panel, 100, resp.  300 experiments), the score
saturates and after a certain point the model does not improve anymore
by increasing the number of modules.  Hence, the optimal number of
modules should be situated around the point where the Bayesian score
starts to level off.  For increasing number of experiments, the
optimal number shifts to the right. This suggests that increasing
amounts of data enable the algorithm to uncover smaller and more
finetuned modules. However, the rightbound shift of the optimum
becomes less pronounced for increasing number of experiments. This
reflects the fact that only a limited number of modules are inherently
present in the true network.

We define the number of modules in the true network as the number of
gene sets having the same set of regulators (taking into account
activator or repressor type). This number is $286$ for the 1000 gene
synthetic network we consider here, among which there are $180$ with
at least $3$ genes and $126$ with at least $5$ genes.  The saturation
behavior of the score curves for 100 and 300 experiments in
Figure~\ref{fig:score_mod} more or less reflects the modularity in the
true network.

\subsection*{Network inference performance}

A more detailed analysis of network inference performance is obtained
by comparing the set of regulator to gene edges in the true
(synthetic) network and in the inferred module network. We use
standard measures such as recall, precision, and $F$-measure (see
Methods).

Figure~\ref{fig:rec_mod} shows the recall as a function of the
number of modules for different numbers of experiments. The location
of the recall maxima seems to agree well with the saturation
points of the corresponding Bayesian score curves
(Figure~\ref{fig:score_mod}). As expected the maximal recall, and
hence the total number of true positives, increases for data sets with
more experiments, saturating between $30$ and $35\%$ for data sets
with $\geq 100$ experiments.

A similar saturation with increasing number of experiments is seen for
the precision curves (Figure \ref{fig:prec_mod} in Additional file 1)
and the $F$-measure curves (Figure \ref{fig:f_mod} in Additional file
1).  Whereas the precision continues to increase with the number of
modules, the $F$-measure saturates, but does so at a higher number of
modules than the Bayesian score.  Taking into account the modular
composition of the true network (see previous section), the Bayesian
score and the recall curves seem to generate better estimates of the
optimal number of modules than the $F$-measure curves.

\begin{figure}[H]
  \centering
  \includegraphics[width=\linewidth]{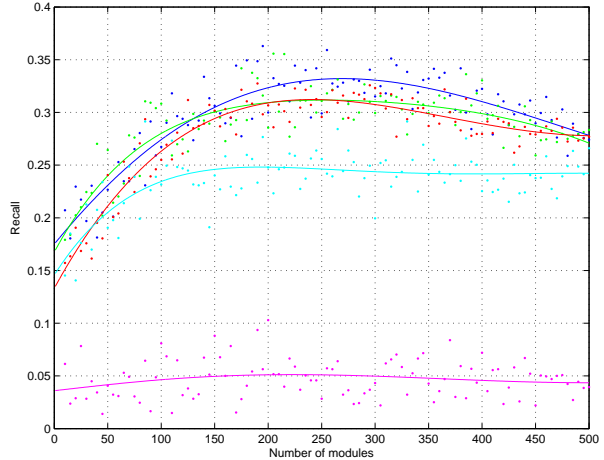}
  \caption{Recall as a function of the number of modules for data
    sets with 10 (magenta), 50 (cyan), 100 (red), 200 (green), and 300
    (blue) experiments.  The curves are least squares fits of the data
    to a linear non-polynomial model of the form $a_0 + \sum_{k=1}^n
    a_k x^{k-1} e^{-x/500}$ with $x$ the number of modules and $n=3$.}
  \label{fig:rec_mod}
\end{figure}

We also investigated whether the inferred regulation programs provide
any information regarding the quality of the regulators. When
analyzing real data, such information could be useful to prioritize
regulators for experimental validation. A first property which we
tried to relate to a regulator's quality is its hierarchical location
in the regulation program. It seems that regulators deeper in the
regulation tree become progressively less relevant.
Figure~\ref{fig:prec_mod_level} illustrates this effect by showing
separately the precisions for the roots of the regulation trees (level
$0$), the children of the roots (level $1$), and the grandchildren
(level $2$) for data sets with 100, 200, and 300 experiments.  The
precisions for the various regulatory levels remain within each others
standard deviation across the tested range of experiments, but the
precision clearly diminishes with increasing levels in the regulation
program. For each data set and inferred module network we created an
additional network where each module is assigned a random regulator
set of the same size as in the inferred network.  The precision for
these random regulation programs is shown in the bottom most curves in
Figure~\ref{fig:prec_mod_level}. For regulation levels beyond level
$2$, the precisions fall in this region of random assignments and they
add almost exclusively false positives (results not shown).  In
general, we can say that the top regulators are far more likely to
represent true regulatory interactions.

\begin{figure}[H]
  \centering
  \includegraphics[width=\linewidth]{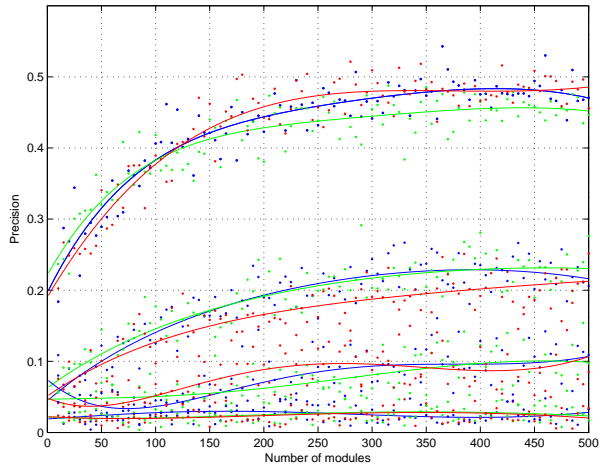}
  \caption{Precision as a function of the number of modules for
    subnetworks generated by regulation tree levels 0 (roots), 1 and
    2, and for random assignments of regulators to regulation tree
    nodes (top to bottom) for data sets with 100 (red), 200 (green)
    and 300 (blue) experiments. The curves are least squares fits of
    the data to a linear non-polynomial model of the form $a_0 +
    \sum_{k=1}^n a_k x^{k-1} e^{-x/500}$ with $x$ the number of
    modules and $n=3$.}
  \label{fig:prec_mod_level}
\end{figure}

An additional layer of information is provided by the regulator
assignment entropies. A low value of the entropy corresponds to a
regulator matching well with a split in the expression pattern of the
regulated module. Hence we expect regulators with low entropy to have
a higher probability to be true regulators. This is illustrated in
Figure~\ref{fig:prec_ent}.  For the data set with 100 experiments and
150 modules, the subnetwork generated by all regulators with an
entropy lower than, e.g., $0.1$ has precision $0.334$, almost twice as
high as the precision of $0.176$ for the whole module network. For the
subnetwork generated by the regulators at the roots of the regulation
trees, the precision increases from $0.42$ to $0.53$ by introducing
the same entropy cut-off.  Other data sets show similar behavior (data
not shown).

\begin{figure}[H]
  \centering
  \includegraphics[width=\linewidth]{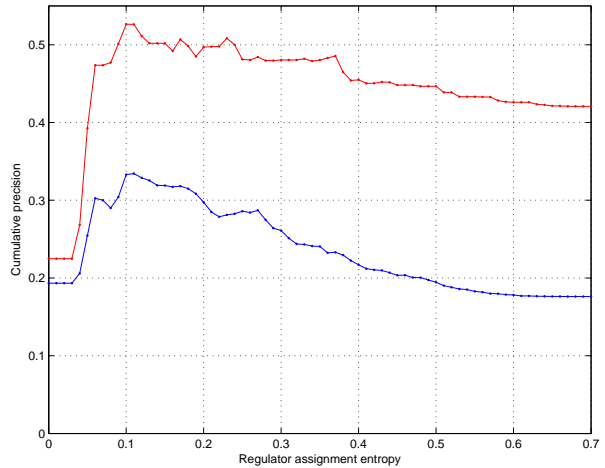}
  \caption{Cumulative distribution of precision as a function of
    regulator entropy for the data set with 100 experiments and 150
    modules: each point at an entropy value $x$ (spaced at 0.01
    intervals) gives the precision of all (blue) or top (red)
    regulators with assignment entropy $\leq x$.}
  \label{fig:prec_ent}
\end{figure}

\subsection*{Performance of \lemone{} versus Genomica}

Next, we compared the performance of \lemone{} and Genomica
\cite{segal2003,genomica}.  Both programs heuristically search for an
optimal module network and are therefore bound to end up at a
(different) local maximum of the Bayesian score. We simulated 10
different data sets with 100 experiments for the same 1000 gene
network as before and inferred a network with $150$ modules
(corresponding to the point where the score function in
Figure~\ref{fig:score_mod} starts to saturate).  The average
precisions are $0.196\pm 0.015$, resp.  $0.155\pm0.013$, and average
recalls $0.255\pm0.016$, resp.  $0.381\pm0.021$, for \lemone{}, resp.
Genomica. The average $F$-measure is $0.222\pm0.015$, resp.
$0.220\pm0.016$. The similarity in performance at the level of the
whole module network, with a bias for higher precision in \lemone{}
and higher recall in Genomica, is further seen in
Figure~\ref{fig:noise}, where we plot recall -- precision pairs for
both programs at different noise levels. For each of the plotted
series, lower noise levels correspond to points in the upper right of
the series plot, and higher noise levels to points in the bottom left,
illustrating a general decrease in performance for more noisy data.

The average module overlap between the module networks generated by
\lemone{} and Genomica is $0.46 \pm 0.02$. Both programs, although
featuring similar performance, attain a different local maximum of the
Bayesian score, and the differences in the corresponding module
networks can be quite substantial.

In general we can say that both module network inference programs
suffer from a high number of false positive edges. When using
\lemone{}, false positives can to some extent be filtered out by
looking only at the highest levels in the regulation tree
(Figure~\ref{fig:prec_mod_level}). To see whether this is also the
case for Genomica, we calculated the recall and precision for the
subnetworks generated by the top regulators alone
(Figure~\ref{fig:noise}). The recall for these subnetworks is
generally lower as they contain far fewer edges than the complete
module network.  For \lemone{} this decrease in recall is
compensated by a large increase in precision. For Genomica the
decrease in recall is bigger, with only a slight increase in
precision.  There is no analogue of the assignment entropy in
Genomica, so we cannot compare the gain in precision by imposing an
entropy cut-off.

One of the major differences in \lemone{} with respect to Genomica is
the fact that the regulatory tree structures learned by \lemone{} are
only dependent on the expression data inside the module, and not on
the expression profiles of potential regulators.

\begin{figure}[H]
 \centering
  \includegraphics[width=\linewidth]{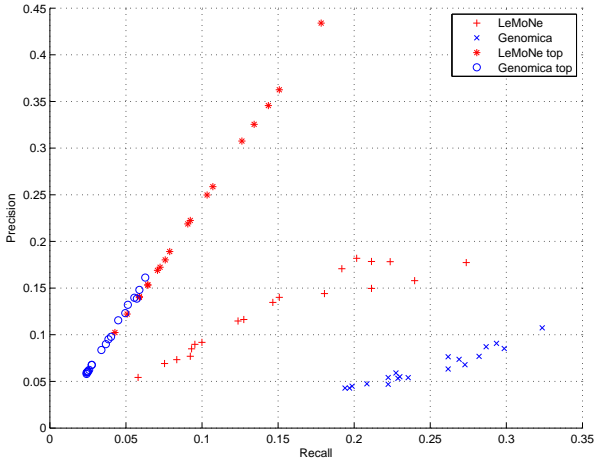}
  \caption{Comparison of heuristic search methods by recall --
    precision pairs for data sets with 100 experiments and different
    noise levels, for the complete module network, and for the
    subnetwork generated by the top regulators in the regulation
    programs.}
  \label{fig:noise}
\end{figure}

We hypothesized that this might allow \lemone{} to better handle
missing or hidden regulators, a situation which might for instance
occur if the true regulator is missing from the list of potential
regulators. In order to test this hypothesis, we simulated 10
different data sets with 100 experiments for the same 1000 gene
network and inferred module networks with 150 modules using both
\lemone{} and Genomica.  In each of the ten runs we randomly left out
20\% of the potential regulators from the regulator list (i.e., we
used 84 instead of 105 potential regulators). The average $F$-measure
of the resulting networks is $0.183\pm0.025$ for \lemone{}, versus
$0.126\pm0.012$ for Genomica.  Compared to the results when taking
into account all 105 potential regulators (see above), the performance
drop for \lemone{} is clearly less pronounced ($17.6\%$) than for
Genomica ($42.7\%$), indicating that \lemone{} is indeed better at
handling missing regulators.

Regarding the speed of \lemone{} versus Genomica, we can say that
\lemone{} is considerably faster for larger data sets. This is mainly
due to the fact that in \lemone{} the regulators need only be assigned
to the regulation programs once, after the final convergence of the
Bayesian score. This saves a considerable amount of time on scanning
possible split values and performing acyclicity checks at each
iteration.  Roughly, \lemone{} and Genomica performed equally in terms
of speed on the \syntren{} data set containing 1000 genes and 100
experiments. On a real data set with 173 experiments \cite{gasch2000},
\lemone{} was about twice as fast as Genomica when limiting the number
of genes to 1000, and ten times faster when considering the whole data
set (2355 genes).

\subsection*{Biological data}

For real biological data sets the underlying regulatory network is
generally not known (indeed, the primary purpose of module network
learning algorithms is precisely to infer the regulatory network) and
hence it is difficult to assess the quality of an inferred network.
This is one of the main reasons why microarray data simulators such as
\syntren{} have to be used to validate the methodology. However, given
the fact that data simulators seldomly capture all aspects of real
biological systems, any results obtained on simulated data should be
approached critically and, where possible, validated on biological
data sets. Here, we investigate to what extent module networks inferred
from real expression data can be validated using structured biological
information.

For \emph{S. cerevisiae}, there is partial information on the
underlying network structure in the form of ChIP-chip data and
promoter motif data \cite{harbison2004}, and more profusely in the
form of GO annotations \cite{ashb00}.  We learned module networks for
budding yeast from an expression compendium containing data for 2355
genes under 173 different stress conditions \cite{gasch2000} (the
Gasch data set) using the same number of modules (50) and the same
list of potential regulators as Segal \emph{et al.} \cite{segal2003}.
We then calculated the $F$-measure between the resulting regulatory
network and the ChIP-chip network of Harbison \emph{et al.}
\cite{harbison2004}, considering in the former network only regulators
that were tested by ChIP-chip. In general, the resulting recall
and precision values are substantially lower than for simulated data
of the same size, namely $0.0195$, resp $0.0218$. When looking at
individual modules, only $13$ out of $50$ regulatory programs feature
at least one regulator that is to some extent confirmed by ChIP-chip
data. In addition, we tried to relate the regulatory program of a
module to the module's gene content in functional terms using GO
annotation. Overall, only $8$ out of $50$ programs possess one or more
regulators belonging to a yeast GOSlim Biological Process category
that is overrepresented in the module (considering only the leaf
categories in the GOSlim hierarchy). Remarkably, only $3$ of these $8$
programs overlap with the $13$ regulatory programs featuring overlap
with the ChIP-chip data. This observation suggests that both data
types can actually be used only to a limited extent to infer the
quality of regulation programs. Indeed, many factors limit the use of
ChIP-chip and GO data as `gold standards'. Both types of data are
noisy and offer incomplete information. For example, Harbison \emph{et
  al.} \cite{harbison2004} mainly profiled transcription factor
binding in rich medium conditions, whereas the Gasch data set contains
primarily stress conditions. The parts of the transcriptional network
that are active under these conditions may substantially differ
\cite{harbison2004, luscombe2004}. Moreover, the expression profile of
a transcription factor is often not directly related to the expression
profile of its targets, for example due to post-translational
regulation of transcription factor activity. As a consequence,
indirect regulators such as upstream signal transducers may feature in
the regulation programs instead of the direct regulators, i.e., the
transcription factors \cite{segal2003}. As for GO, many regulators
appear not to be annotated to the GO Biological Process categories of
their target genes. Taking these factors into account, the limited
overlap with the available ChIP-chip and GO data does not necessarily
reflect the quality of the inferred regulatory programs.

On the contrary, we established that the regulatory programs do in
fact contain a considerable amount of relevant and potentially
valuable information. Indeed, by manually investigating individual
modules in more detail, we could in many cases qualitatively relate
the regulators to the module's gene content. For example, the module
shown in Figure \ref{fig:gasch_module_11} is enriched in a.o. genes
involved in the main pathways of carbohydrate metabolism
($P=1.0596$E$-4$), energy derivation by oxidation of organic compounds
($P=1.2046$E$-4$) and alcohol biosynthesis ($P=1.3185$E$-2$). None of
the $5$ regulators of this module could be related to the module's
gene content based on ChIP-chip or GO information. However, based on
their description in the {\it Saccharomyces} Genome Database (SGD)
\cite{sgd}, all $5$ regulators could be linked to glucose sensing or
the response to (glucose) starvation, processes that can arguably
influence the expression of carbohydrate metabolism genes.

\begin{figure}[H]
 \centering
  \includegraphics[width=\linewidth]{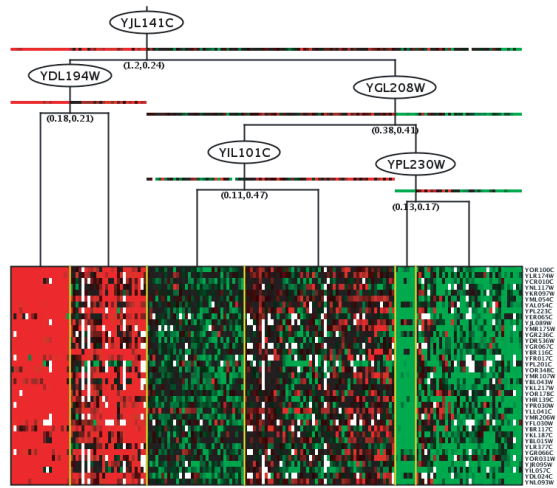}
  \caption{Sample module learned from the Gasch data set
    \cite{gasch2000}. Red and green hues indicate upregulation resp.
    downregulation. The pairs $(x,y)$ under each split in the
    regulation tree represent the Bayesian score gain over the split,
    normalized on the number of genes in the complete network ($x$),
    and the regulator assignment entropy ($y$).}
  \label{fig:gasch_module_11}
\end{figure}

However, one must keep in mind that it remains impossible to infer
complete and accurate regulatory networks from gene expression data
alone. Expression data only provides information on one regulatory
level, namely the transcriptional level. Information on
(post-)translational regulation is lacking. The current
expression-based module network algorithms (e.g. \cite{segal2003},
this study) try to remedy this problem by including signal transducers
in the list of potential regulators in addition to transcription
factors, in the hope to capture some of this non-transcriptional
regulation from the expression profiles of key signal transducers.
However, this trick can only be expected to uncover a fraction of such
non-transcriptional regulatory interactions, and moreover the direct
targets of these regulatory interactions are not identified. A
potential remedy for this shortcoming would be to include other types
of data, such as data on protein expression levels and protein
phosphorylation, in the module network learning framework.
Unfortunately, such data are not yet available on a large scale.

In summary, our results indicate that structured biological
information such as ChIP-chip data or GO can not (yet) be used to
measure the performance of module network algorithms in an automated
way. This is a strong argument for using data simulators such as
\syntren{} for the purpose of developing, testing and improving such
algorithms.

\section*{Conclusions}

We developed a module network learning algorithm called \lemone{} and
tested its performance on simulated expression data sets generated by
\syntren{} \cite{vandenbulcke2006b}. We found that the Bayesian score
can be used to infer the optimal number of modules, and that the
inference performance increases as a function of the number of
simulated experiments but saturates well below 1.

We also used \syntren{} data to assess the effects of the
methodological changes we made in \lemone{} with respect to the
original methods used in Genomica \cite{segal2003}. Overall,
application of Genomica and \lemone{} to various simulated data sets
gave comparable results, with a bias towards higher recall for
Genomica and higher precision for \lemone{}. However, \lemone{} offers
some advantages over the original framework of Segal \emph{et al.}
\cite{segal2003}, one of them being that the learning process is
considerably faster.  Another advantage of \lemone{} is the fact that
the algorithm `lets the data decide' when learning the regulatory tree
structure. The partitioning of expression data inside a module is not
dependent on the expression profiles of the potential regulators, but
only on the module data itself. As a consequence, the assignment of
`bad' regulators (in terms of assignment entropy) to `good' module
splits (in terms of Bayesian score) might suggest missing or hidden
regulators. This situation might occur if the true regulator is
missing from the list of potential regulators, or if the expression of
the targets cannot be related directly to the expression of the
regulator, e.g., due to posttranslational regulation of the
regulator's activity.  We have also shown that filtering the module
network by the location of regulators in the regulation program or by
introducing an entropy cut-off improves the inference performance.
When inferring regulatory programs from real data, these criteria may
prove useful to prioritize regulators for experimental validation.

\enlargethispage{\baselineskip}

Finally, we explored the extent to which module networks inferred from
real expression data could be validated using structured biological
information. For that purpose, we learned module networks from a
microarray compendium of stress experiments on budding yeast
\cite{gasch2000}. We found that the resulting regulatory programs
overlapped only marginally with the available ChIP-chip data and GO
information. However, more detailed manual analysis uncovered that the
learned regulation programs are nevertheless biologically relevant,
suggesting that an automated assessment of the performance of module
network algorithms using structured biological information such as
ChIP-chip data or GO is ineffective. This underscores the importance
of using data simulators such as \syntren{} for the purpose of testing
and improving module network learning algorithms.


\section*{Authors contributions}
T.M. and S.M. designed the study, developed software, analyzed the
data and wrote the paper. E.B. and A.J. designed the study, developed
software and analyzed the data. Y.S. designed the study and developed
software. T.V.d.B. and K.V.L. developed software. P.v.R., M.K., K.M.
and Y.V.d.P. designed the study and supervised the project.

\section*{Acknowledgements}
\ifthenelse{\boolean{publ}}{\small}{} We thank Eran Segal for
explanation about the Genomica algorithm, and Gary Bader and Ruth
Isserlin for refactoring the BiNGO code which allowed its
incorporation into LeMoNe.  T.M. and S.M.  are Postdoctoral Fellows of
the Research Foundation Flanders (Belgium), A.J. is supported by an
Early Stage Marie Curie fellowship.  This work is partially supported
by: IWT projects: GBOU-SQUAD-20160; Research Council KULeuven:
GOA-Ambiorics, CoE EF/05/007 SymBioSys; FWO projects: G.0413.03, and
G.0241.04.




\ifthenelse{\boolean{publ}}{\end{multicols}}{}

\newpage


\setcounter{figure}{0}
\renewcommand{\thefigure}{S\arabic{figure}}
\setcounter{equation}{0}
\renewcommand{\theequation}{S\arabic{equation}}

\begin{flushleft}
  {\sffamily{\raggedright{\noindent\LARGE \bfseries Additional file 1: Supplementary
        information\par}}}%
  \end{flushleft}

\ifthenelse{\boolean{publ}}{\begin{multicols}{2}}{}

\section*{Precision and F-measure as a function of the number of
    modules and experiments}

\ifthenelse{\boolean{publ}}{\end{multicols}}{}

\ifthenelse{\boolean{publ}}{\begin{multicols}{2}}{}

\begin{figure}[H]
  \centering
  \includegraphics[width=\linewidth]{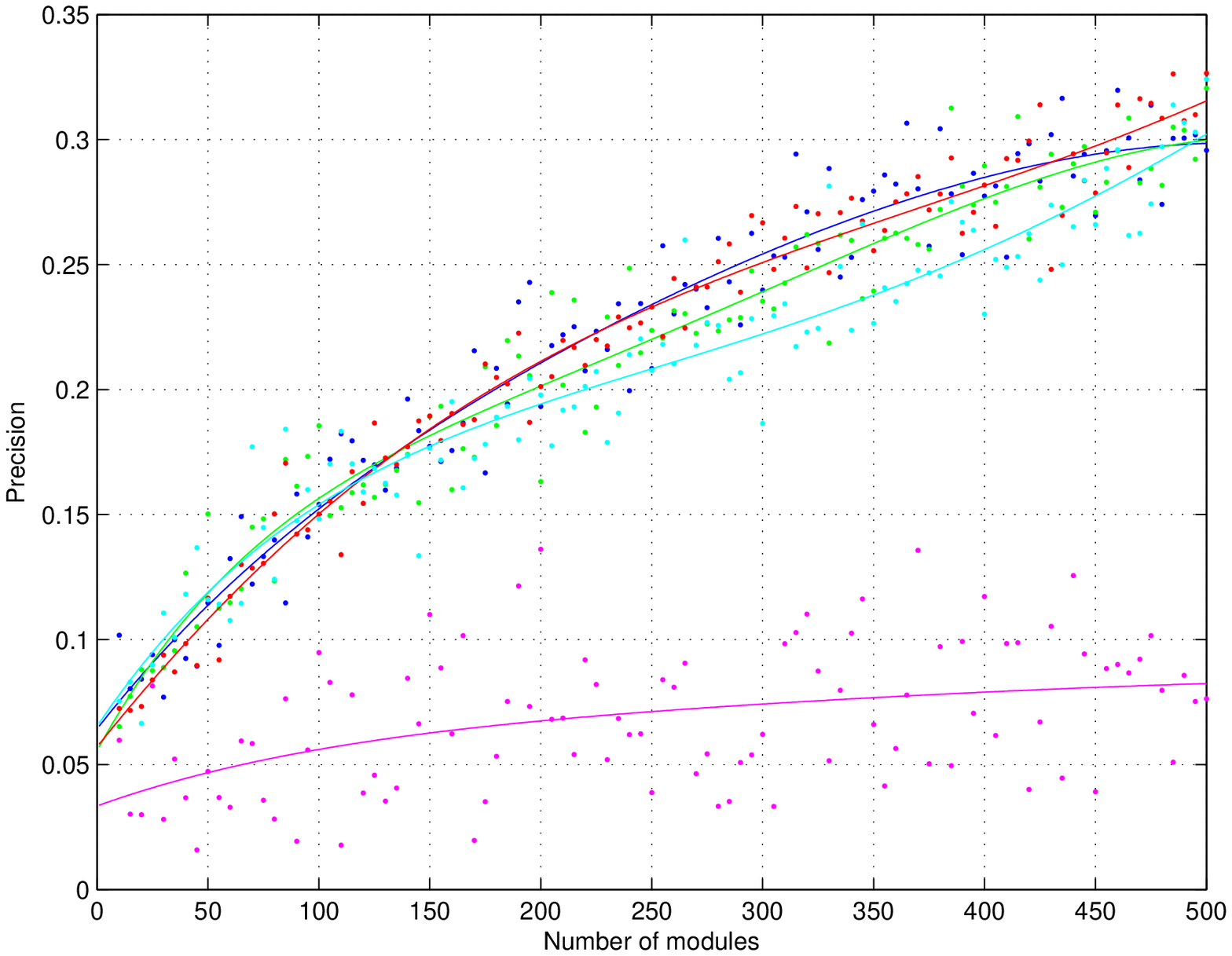}
  \caption{Precision as a function of the number of modules for data sets with
   10 (magenta), 50 (cyan), 100 (red), 200 (green), and 300 (blue)
   experiments.  The curves are least squares fits of the data to a
   linear non-polynomial model of the form $a_0 + \sum_{k=1}^n a_k
   x^{k-1} e^{-x/500}$ with $x$ the number of modules and $n=3$.}
  \label{fig:prec_mod}
\end{figure}

\begin{figure}[H]
  \centering
  \includegraphics[width=\linewidth]{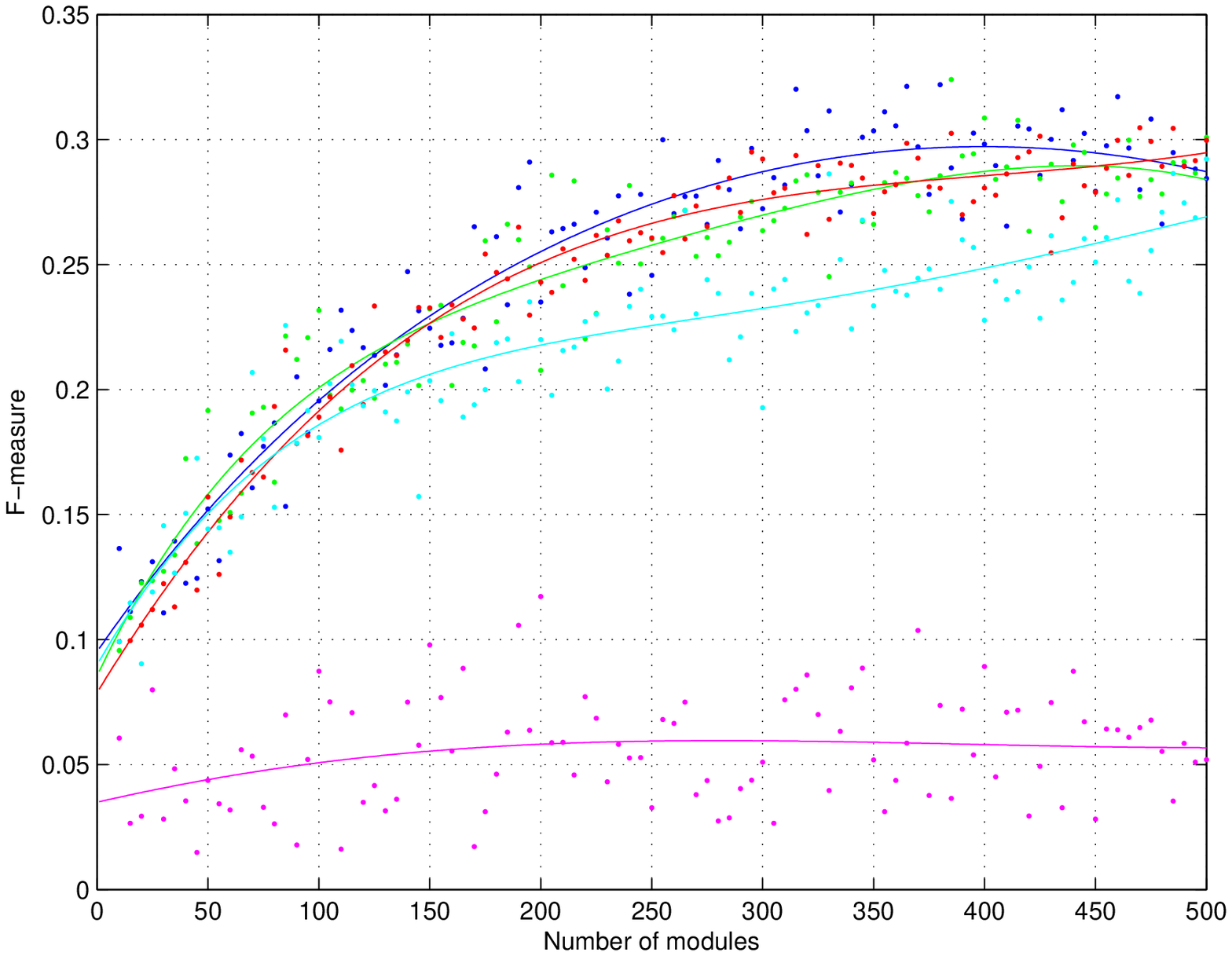}
  \caption{$F$-measure as a function of the number of modules for data
    sets with 10 (magenta), 50 (cyan), 100 (red), 200 (green), and 300
    (blue) experiments.  The curves are least squares fits of the data
    to a linear non-polynomial model of the form $a_0 + \sum_{k=1}^n
    a_k x^{k-1} e^{-x/500}$ with $x$ the number of modules and $n=3$.}
  \label{fig:f_mod}
\end{figure}

\ifthenelse{\boolean{publ}}{\end{multicols}}{}

\ifthenelse{\boolean{publ}}{\begin{multicols}{2}}{}

\section*{Module networks Bayesian score}

We use the same Bayesian score as in the original module networks
formalism \cite{segal2003,segal2005}. The data likelihood is given by
evaluating the module network joint probability distribution (eq.~(2))
on the data set, assuming independent experiments,
\begin{equation*}
  {\cal L} = \prod_{m=1}^M \prod_{k=1}^K \prod_{i\in\A_k} p_k 
  \bigl(x_{i,m} \mid \{x_{j,m}\colon   j\in\Pi_k\}\bigr),
\end{equation*}
where $x_{i,m}$ is the $\log$-normalized expression value of gene $i$
in experiment $m$.

The Bayesian score is obtained by taking the $\log$ of the marginal
probability of the data likelihood over the parameters of the normal
distributions at the leaves of the regression trees with a
normal-gamma prior. It decomposes as a sum of leaf scores of the
different modules:
\begin{align}
  \mathbf{S} &= \sum_k \mathbf{S}_k = \sum_k\sum_\ell S_k(\E_\ell)\nonumber\\
  S_k(\E_\ell) &= \log \iintsp d\mu d\tau\, p(\mu,\tau)
  \prod_{m\in\E_\ell}\prod_{i\in\A_k}p_{\mu,\tau}(x_{i,m}),
  \label{eq:S1}
\end{align}
where $k$ runs over the set of modules and $\ell$ runs over the set of
leaves of the regression tree of module $k$; $\E_\ell$ denotes the
experiments that end up at leaf $\ell$ after traversing the regression
tree and $\A_k$ denotes the genes assigned to module $k$;
$p(\mu,\tau)$ is a normal-gamma distribution over the mean $\mu$ and
precision $\tau$ of the normal distribution $p_{\mu,\tau}$, i.e.,
$p(\mu,\tau)=p(\mu\mid\tau)p(\tau)$ where
$p(\tau)\sim\Gamma(\alpha_0,\beta_0)$ and $p(\mu\mid\tau)\sim{\cal
  N}(\mu_0, (\lambda_0\tau)^{-1})$:

\begin{align}
  p(\tau) &= \frac{\beta_0^{\alpha_0}}{\Gamma(\alpha_0)}
  \tau^{\alpha_0-1} e^{-\beta_0\tau} \label{eq:S2}\\
  p(\mu\mid\tau) &= \bigl(\frac{\lambda_0\tau}{2\pi}\bigr)^{1/2}
  e^{-\frac{\lambda_0\tau}2 (\mu-\mu_0)^2}\label{eq:S3}\\
  p_{\mu,\tau}(x_{i,m}) &= \bigl(\frac{\tau}{2\pi}\bigr)^{1/2}
  e^{-\frac\tau2(x_{i,m}-\mu)^2}\label{eq:S4}
\end{align}
with $\alpha_0,\beta_0,\lambda_0 > 0$ and $-\infty<\mu_0<\infty$.

Insertion of eqs. (\ref{eq:S2})--(\ref{eq:S4}) into eq. (\ref{eq:S1})
leads to an integral that can be solved explicitly as a function of
the sufficient statistics 
\begin{equation*}
  R_q^{(\ell)} = \sum_{m\in\E_\ell} \sum_{i\in\A_k} x_{i,m}^q\,,\; q=0,1,2.
\end{equation*}
of the leaves of the regression tree. The result is
\begin{align}
  S_k(\E_\ell) =& -\tfrac12 R_0^{(\ell)}\log(2\pi) + \tfrac12
  \log\bigl(\frac{\lambda_0}{\lambda_0 + R_0^{(\ell)}}\bigr) \nonumber \\
  & - \log\Gamma(\alpha_0) + \log\Gamma(\alpha_0
  + \tfrac12 R_0^{(\ell)}) \nonumber\\
  &+ \alpha_0\log\beta_0 -(\alpha_0 + \tfrac12
  R_0^{(\ell)})\log\beta_1 \label{eq:S9}
\end{align}
where
\begin{multline*}
  \beta_1 = \beta_0 + \frac12\Bigl[ R_2^{(\ell)} -
  \frac{(R_1^{(\ell)})^2}{R_0^{(\ell)}} \Bigr] \\
  + \frac{\lambda_0 \bigl( R_1^{(\ell)} - \mu_0 R_0^{(\ell)}
    \bigr)^2}{2(\lambda_0 + R_0^{(\ell)})R_0^{(\ell)}}.
\end{multline*}

\section*{Learning module regulation programs}

The pseudocode for the regulation program learning algorithm is given
in Figure \ref{fig:code}. In its simplest form, the merge score for
two trees $T_{\alpha_1}$ and $T_{\alpha_2}$ considers only the gain in
Bayesian score that is obtained by merging two sets into one:
\begin{equation}\label{eq:S6}
  r_{\alpha_1,\alpha_2} = S_k(\E_{\alpha_1}\cup\E_{\alpha_2}) 
  - S_k(\E_{\alpha_1}) - S_k(\E_{\alpha_2}).
\end{equation}
In our computations we used a merge score which is slightly more
complicated and takes into account the whole substructure of the tree
below $T_{\alpha_1}$ and $T_{\alpha_2}$. 

Let $T_\alpha$ be a tree with children $T_{\alpha_1}$ and
$T_{\alpha_2}$, and define recursively
\begin{align*}
  Z_\alpha &= e^{S_k(\E_\alpha)} + Z_{\alpha_1}Z_{\alpha_2}
\intertext{with initial condition}
  Z_m &= e^{S_k(\{m\})}
\end{align*}
for the trivial tree with one experiment $m$ and no children. The new
merge score is then defined as
\begin{equation}\label{eq:S7}
  r_{\alpha_1,\alpha_2} = S_k(\E_{\alpha_1}\cup\E_{\alpha_2}) - \ln Z_{\alpha_1}
  - \ln Z_{\alpha_2}.
\end{equation}

A binary tree $T_\alpha$ generates a nested set of partitions
$\P_\alpha$ (we write this as $\P_\alpha\sim T_\alpha$) of its
experiment set $\E_\alpha$ and to each such partition corresponds a
score
\begin{equation*}
  \mathbf{S}_k(\P_\alpha) = \sum_i S_k(\E_i)
\end{equation*}
where $\E_i$ are the subsets of $\E_\alpha$ forming the partition
$\P_\alpha$. Since a partition generated by $T_\alpha$ is either the
singleton partition $\P_\alpha=\{\E_\alpha\}$, or a combination of a
partition generated by $T_{\alpha_1}$ with a partition generated by
$T_{\alpha_2}$, we get immediately
\begin{equation*}
  Z_\alpha = \sum_{\P_\alpha\sim T_\alpha} e^{\mathbf{S}_k(\P_\alpha)},
\end{equation*}
or
\begin{equation}\label{eq:S10}
  \ln Z_\alpha = S_k(\E_\alpha) + \ln\Bigl( 1 + \sum_{\substack{\P_\alpha\sim T_\alpha\\
      \P_\alpha\neq \{\E_\alpha\}}}
  e^{\mathbf{S}_k(\P_\alpha)-S_k(\E_\alpha)} \Bigr)
\end{equation}
We conclude that the merge score (\ref{eq:S7}) contains the score
difference (\ref{eq:S6}) as well as other terms defined by the
structure of the subtrees $T_{\alpha_1}$ and $T_{\alpha_2}$. If two
pairs of trees give the same score difference (\ref{eq:S6}), the merge
score (\ref{eq:S7}) will typically favor to merge the pair with the
smallest substructure first (as the number of terms in the summation
in (\ref{eq:S10}) is smaller). Hence, using (\ref{eq:S7}) instead of
(\ref{eq:S6}) leads to more balanced trees.

Since we are building the tree from the bottom up, computing the
partition sums $Z_\alpha$ is done along the way, and using the merge
score (\ref{eq:S7}) instead of (\ref{eq:S6}) comes at no computational
cost. The whole process of constructing the tree with a merge score
depending on all the partitions generated by the subtrees is very
similar to the Bayesian hierarchical clustering method of
\cite{heller2005}.

\section*{Regulator assignment in the presence of missing values}

In real data there are often missing values, so for some experiments
we do not know if a regulator is above or below a given split value.
Using the non-missing values to define the sets $\R_1$ and $\R_2$, we
compute $q_i$ as before.  Since regulators with a lot of missing
values lead to more uncertainty, we penalize those by moving the
conditional probabilities $q_i$ closer to the maximum uncertainty
value of $\tfrac12$ by defining
\begin{equation*}
  q_i' = (1-\tfrac{|\R_3|}{|\E_\alpha|})\,(q_i-\tfrac12) + \tfrac12,
\end{equation*}
where
\begin{equation*}
  \R_3 = \{m\in\E_\alpha\colon x_{r,m} \text{ is missing} \}.
\end{equation*}
Note that when there are no missing values, $q_i'=q_i$, and in the
extreme case where there are only missing values, $q_i'=\tfrac12$.
For the probability distribution of $R$, we distribute the missing
values proportionally over $1$ and $2$,
\begin{equation*}
  p_i' = \frac{|\R_i|+ \frac{|\R_i|}{|\R_1|+|\R_2|}|\R_3|}{|\E_\alpha|}.
\end{equation*}
such that $p_1'+p_2'$ still sums up to $1$. We now minimize the
conditional entropy 
\begin{equation*}
  H(E\mid R) =  p_1' h(q_1')  + p_2' h(q_2')
\end{equation*}
corresponding to these modified probability distributions.  For the
sufficient statistics of the leaves of the module, we simply ignore
the missing values as there are typically more than enough combined
data points for a reliable computation of those statistics.

The complete regulator assignment algorithm is given in Supplementary
Figure~\ref{fig:code2}

\ifthenelse{\boolean{publ}}{\end{multicols}}{}

\ifthenelse{\boolean{publ}}{\begin{multicols}{2}}{}

\begin{figure}[H]
 \centering
 
 \begin{algorithm}[H]
   \tcc{Find hierarchical tree}
   \KwIn{A list \textbf{treeList} of trivial trees representing single
     experiments.}
   \While{\textbf{treeList} has more than 1 element}{
     compute $r_{\alpha_1,\alpha_2}$ for each pair of trees in 
     \textbf{treeList}\;
     construct the joined tree $T_\alpha=T_{\alpha_1}\cup T_{\alpha_2}$ 
     for the pair with highest $r_{\alpha_1,\alpha_2}$\;
     add $T_\alpha$ to \textbf{treeList} and remove $T_{\alpha_1}$,
     $T_{\alpha_2}$\;
   }
   \KwOut{A single tree $T_0$ representing all experiments.}
   \vspace*{\baselineskip}

   \tcc{Find optimal regulation program leaves}
   \SetKwFunction{KwFn}{testScore}
   \KwFn{\textbf{$T_0$.root}}\;
   \vspace*{\baselineskip}

   \tcc{Recursive procedure to cut the hierarchical tree}
   \SetKwBlock{Begin}{Begin testScore(node)}{End}
   \Begin{ 
     \eIf{$S_k$(\textbf{node}) $<$ $S_k$(\textbf{node.leftChild}) +
       $S_k$(\textbf{node.rightChild})}{
       testScore(\textbf{node.leftChild})\;
       testScore(\textbf{node.rightChild})\;
     }  {
       cut tree below \textbf{node};
     }
   }
 \end{algorithm}
 \caption{Pseudocode for the regulation program learning method}
 \label{fig:code}
\end{figure}

\begin{figure}[H]
 \centering
 
 \begin{algorithm}[H]
  \tcc{Assign regulators separately for different regulation tree levels.}
  \tcc{Level is the distance from a node to the root.}
  \For{each level $l$}{
    create a list \textbf{nodeList} with all nodes at level $l$ in 
    the trees of all modules\;
    \While{\textbf{nodeList} is not empty}{
      \For{each \textbf{node} in \textbf{nodeList}}{
        \For{each regulator \textbf{r} in the set of potential regulators 
          that do not break acyclicity}{
          compute the entropy for assigning \textbf{r} to \textbf{node}\;
        }
      }
      find the node-regulator pair (\textbf{bestNode}, \textbf{bestR}) with 
      least entropy\;
      assign \textbf{bestR} to \textbf{bestNode}\;
      remove \textbf{bestNode} from \textbf{nodeList}\;
    }
  }
 \end{algorithm}
 \caption{Pseudocode for the regulation program learning method}
 \label{fig:code2}
\end{figure}


\ifthenelse{\boolean{publ}}{\end{multicols}}{}

\end{bmcformat}
\end{document}